# Nonvolatile Silicon Photonic MEMS Switch Based on Centrally-Clamped Stepped Bistable Mechanical Beams

QIAN MA[1], YINPENG HU[1], YE LU[1], YUNZHI LIU[1], HUAN LI[1,2,3*], DAOXIN DAI[1,2,3]

[1]State Key Laboratory for Modern Optical Instrumentation, Center for Optical & Electromagnetic Research, College of Optical Science and Engineering, International Research Center for Advanced Photonics, Zhejiang University, Zijingang Campus, Hangzhou 310058, China
[2]Ningbo Research Institute, Zhejiang University, Ningbo 315100, China
[3]Jiaxing Key Laboratory of Photonic Sensing & Intelligent Imaging, Intelligent Optics & Photonics Research Center, Jiaxing Research Institute, Zhejiang University, Jiaxing 314000, China
*lihuan20@zju.edu.cn

**Abstract:** High-performance photonic switches are essential for large-scale optical routing for AI large models and Internet of things. Realizing nonvolatility can further reduce power consumption and expand application scenarios. We propose a nonvolatile 2×2 silicon photonic micro-electromechanical system (MEMS) switch compatible with standard silicon photonic foundry processes. The switch employs electrostatic comb actuator to change the air gap of the compact horizontal adiabatic coupler and achieves nonvolatility with centrally-clamped stepped bistable mechanical beams. The photonic switch features a 10s μs-scale switching speed and a 10s fJ-scale simulated switching energy within a 100×100 μm$^2$ footprint, with ≤12 V driving voltages. This 2×2 switch can be used in a variety of topologies for large-scale photonic switches, and its nonvolatility can potentially support future photonic FPGA designs.

## 1. Introduction

To address the growing demand for large-scale optical routing for AI large models and Internet of things, optical/photonic switches have been increasingly deployed in high-performance data centers and computing centers. A variety of optical/photonic switches based on different principles and structures have been developed for diverse application scenarios [1]. Among these, silicon photonic MEMS switches stand out due to their low power consumption, low excess loss, and low crosstalk [2]. These characteristics are achieved by manipulating mode coupling and propagation through mechanical motion, with the power consumed by electrostatic actuation in the steady state being nearly zero [3,4]. Furthermore, these switches can be fabricated using the standard silicon photonic foundry processes, which not only reduces fabrication costs but also opens up the possibility of large-scale photonic integration.

A 1×2 silicon photonic MEMS switch with vertical adiabatic couplers [5] demonstrates excellent scalability [6] (240×240) and has been utilized for large-scale silicon photonic LiDAR [7] (128×128). The vertical adiabatic coupler for a single photonic switch incurs an ON-state switching loss of <0.45 dB. However, this design requires an additional silicon layer, which is incompatible with most standard silicon photonic foundry processes currently available. Alternatively, a 1×2 silicon photonic MEMS switch using lateral adiabatic couplers has been designed [8]. Each photonic switch cell employs two 50-μm long lateral adiabatic couplers and achieves an insertion loss of <1 dB. However, 1×2 photonic switches are not applicable to many topologies other than Cross-Bar (e.g., Benes and PI-Loss). Therefore, 2×2 silicon photonic MEMS switches based on split waveguide crossings have been designed to achieve low excess loss and high extinction ratios over large bandwidths of 1400-1700 nm [9,10].

However, all the volatile switches above require continuous drive voltage to maintain their switching states. Even though the power consumption of the electrostatically actuated switches is nearly zero in the steady state, the power consumption of the voltage supply to generate the drive voltage may already be much greater than that of the switches, which beats the purpose of low power consumption. To completely eliminate the power consumption of the driving circuit itself, nonvolatility should be introduced for the photonic switches, which may also pave the way for further realization of photonic field programmable gate array (FPGA) based on silicon photonic circuits [11], expanding the application scenarios. A nonvolatile 1×2 silicon photonic MEMS switch with vertical adiabatic couplers has been designed [12], which employs bistable mechanism of buckling beams based on residual stress relief. However, only the mechanical bistability of the buckling beams has been experimentally verified, while the photonic switching has not yet been demonstrated. Another structure to realize bistability is cosine-shaped centrally-clamped bistable mechanical beams, which were analyzed in detail [13]. However, if such beams are scaled from 100 μm-scale down to 10 μm-scale to fit the size of photonic switches, the significantly decreased forces required for switching make their second stable states no longer stable enough, which compromises the nonvolatility of the whole structure. To achieve better nonvolatility, replacing the uniform inclined beam with a centrally-widened stepped beam in the bistable triangular unit has been investigated [14]. Such approach points the way to the design of high-performance bistable mechanical beams.

In this paper, we propose and implement a nonvolatile 2×2 silicon photonic MEMS switch compatible with standard silicon photonic foundry processes. Fabricated on 220-nm SOI, this switch employs a bidirectional electrostatic comb actuator to drive a compact horizontal adiabatic coupler (HAC). This HAC couples or decouples on a single-layer silicon structure to achieve 2×2 photonic switching operation. To achieve the overall nonvolatility of the photonic switch, we design a centrally-clamped stepped (CCS) bistable mechanical beam by incorporating rigid widened straight beams into flexible cosine-shaped beams. The overall nonvolatility does not rely on residual stress relief after releasing from the buried oxide (BOX). This photonic switch requires 12 V and 9 V drive voltages for switching to OFF and ON states, respectively, and there is no need to maintain the drive voltage after switching is completed. For high optical performance in the limited footprint, inspired by the ideas of fast quasi-adiabatic dynamics (FAQUAD) [15–18], we designed a compact HAC with 70-μm length by optimizing the empirical expression of waveguide width variation through particle swarm optimization (PSO) and obtained a simulated extinction ratio of >20 dB in the bandwidth of 1520-1578 nm. One representative device has been measured with excellent photonic performance across a large bandwidth. In the bandwidth of 1520-1610 nm, the switch has an excess loss of <0.49 dB and crosstalk of <–12.94 dB in ON state, and an excess loss of <0.43 dB and crosstalk of <–22.15 dB in OFF state. At 1569 nm, the switch has a minimum crosstalk of –32.65 dB in ON state. The photonic switch has a switching speed in the 10s μs-scale, a footprint of only 100×100 μm$^2$, and a simulated switching energy of 65 fJ and 46 fJ for switching to OFF and ON states, respectively. The photonic switch can operate up to 20 kHz over 5×10$^7$ cycles without observable performance degradation. By properly managing the residual stress and improving the overlay accuracy, it is expected to fabricate photonic switches with better photonic and mechanical performance. The CCS beam design can be adapted for compatibility with standard silicon photonic foundry processes of 130 nm or even 180 nm, providing a highly feasible solution for mass production. This 2×2 switch can be used in a variety of topologies for large-scale photonic switches, and its nonvolatility can potentially support future photonic FPGA designs.

## 2. Design and operation principles

The 2×2 silicon photonic MEMS switch proposed in this paper primarily consists of three components: CCS beams, an electrostatic comb actuator, and a compact HAC with adjustable gap, as depicted in Fig. 1(a). The MEMS structure, which is suspended in the air after release, can be mechanically displaced by the electrostatic comb actuator. As illustrated in Fig. 1(b), the initial state is designed to be ON state. Upon release of the MEMS structure, due to the relief of the residual compressive stress in the top-layer of SOI, the CCS beams buckle to their first stable states, pushing the movable waveguide of the HAC to move upward (+$y$-direction) toward the fixed waveguide and stop at a position determined by the mechanical stopper. In this state, the HAC is in the coupling state, and the input light will be coupled into the waveguide on the opposite side. When a voltage is applied to the lower fixed comb of the electrostatic comb actuator, the electrostatic force pulls the suspended structure downward (–$y$-direction). The air gap of the HAC then becomes large enough such that the HAC decouples, and almost no light will be coupled into the waveguide on the opposite side, switching it to OFF state, as shown in Fig. 1(c). Meanwhile, CCS beams switch to the second stable states and then maintain the current state after the voltage is removed, realizing the nonvolatility of the photonic switch. After applying a voltage to the upper fixed combs, the structure undergoes the opposite process to return to ON state.

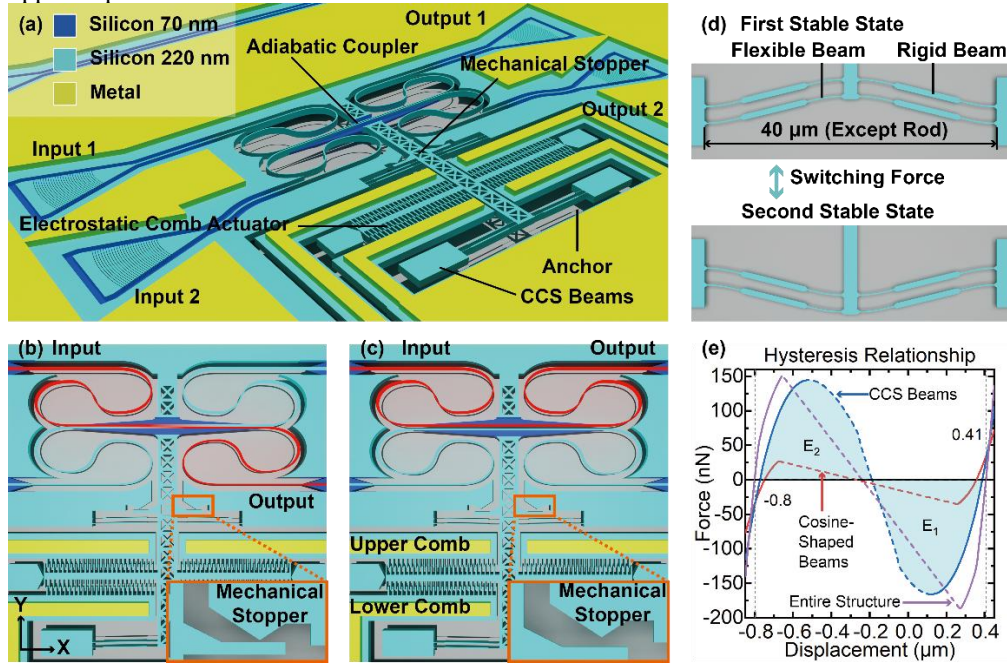

Fig. 1. (a) Full view of the proposed nonvolatile silicon photonic MEMS switch. (b) and (c) ON state and OFF state, respectively. Red waveguide is the path of the transmitted light. (d) State switching process of CCS beams. (e) The force-displacement hysteresis relationship of cosine-shaped beams, CCS beams, and the entire structure. The origin of the horizontal axis is the initial position before releasing the mechanical structures. The CCS beams show a stronger bistability and higher switching forces. The area $E_1$ represents the energy required to switch from the first to the second stable state, while the area $E_2$ represents the energy required for the opposite switching process, for one set of CCS beams. The entire mechanical structure includes two sets of CCS beams, and the hysteresis relationship is influenced by the suspended waveguides.

## 2.1 Centrally-clamped stepped (CCS) bistable mechanical beams

The bistable mechanical beams will switch from one stable state to another when the deformation due to forces in $y$-direction exceeds certain critical values, and the forces needed here are defined as switching forces. For a conventional cosine-shaped beam [13], the switching force from the second stable state to the first one (defined as the switching-ON force) is significantly weaker than that of the reversed switching process (defined as the switching-OFF force). Since the switching-OFF force is limited by drive voltages, the switching-ON force will thus oftentimes be too weak, such that the second stable state will not be stable enough. Therefore, to achieve stronger bistability, the switching-ON force of the bistable beams is preferably required to be as strong as the switching-OFF force. For stronger bistability and higher switching force in the limited footprint, we design CCS bistable mechanical beams consisting of flexible cosine-shaped beams and rigid straight beams, as shown in Fig. 1(d). The flexible cosine-shaped beams and rigid straight beams are designed to be 0.1 μm and 0.4 μm wide, respectively, with smooth transitions (continuous first derivative) between the two types of beams and a total span of 40 μm, and the ratio of their respective lengths in x-direction is 3:7. The Taper transitions are used to connect the flexible cosine-shaped beams, the rigid straight beams, the shuttle beam, and the anchors on both sides to avoid strain concentration during motion. The maximum offset in *y*-direction of the CCS beams from their fixed ends is initially designed to be 0.2 μm. Once the CCS beams are suspended, the relief of the residual stress in the top layer of the SOI wafer, which leads to an estimated pre-strain of ~$4.1 \times 10^{-4}$, will result in a maximum displacement of ~1.2 μm in y-direction between the two stable states, sufficient to generate a large enough air gap of the HAC. The initial offset is sufficiently large to prevents buckling in z-direction. Simulated by finite element method (FEM) software, the force-displacement hysteresis relationship of CCS beams is shown in Fig. 1(e). The ratio of the switching-ON force to the switching-OFF force is 0.87, demonstrating that CCS beams have stronger bistability compared to their cosine-shaped counterparts with the same span and initial offset, for which the ratio is 0.76. The higher switching forces also make the bistability of the CCS beams more robust to any additional mechanical connections, such as the suspended waveguides connected with the rest of the photonic circuits. The theoretical minimum switching energy for one set of CCS beams (Fig. 1(d)) can be estimated to be 64 fJ ($E_1$) and 58 fJ ($E_2$) for switching to the second and first stable states, respectively, by integrating the corresponding hysteresis curves, as shown in Fig. 1(e).

*2.2 Electrostatic comb actuator*

The electrostatic comb actuator consists of a centrally suspended movable comb and fixed combs above (+*y*-direction) and below (–*y*-direction). The fixed combs are connected to respective drive electrodes. Meanwhile, the movable comb is connected to the peripheral silicon layer through waveguides and CCS beams to always be grounded. When voltage is applied to a drive electrode, the connected fixed comb teeth attract the movable comb teeth, causing the latter to move in the former's direction due to the electrostatic force. All comb handles have a total of 100 pairs of comb teeth arranged periodically. Each tooth is designed to be 3 μm long and 0.4 μm wide, with an air gap of 1 μm in-between. The electrostatic force can be calculated using Equation (1):

$$F_e = \frac{1}{2}U^2 \frac{\partial C}{\partial y} \tag{1}$$

where *U* is the voltage applied to the electrostatic comb actuator, *C* is the capacitance between the electrostatic combs, and *y* is the displacement in the *y*-direction. Simulations using FEM software indicate that each pair of comb teeth can generate ~6.8 nN of electrostatic force at 20 V, sufficient for switching between the two stable states of CCS beams.

*2.3 Adiabatic coupler*

The gap-adjustable compact horizontal adiabatic coupler (HAC) comprises a fixed waveguide and a movable waveguide connected to the shuttle beams, as depicted in Fig. 1(a). The two waveguides are centrosymmetric. Shallowly etched trapezoidal slabs are used for simultaneous mechanical connections and optical mode confinement. They are etched to a depth of 150 nm and a width of ~2 μm to prevent evanescent wave leakage, while also enhancing the rigidity of the coupling region and reducing loss due to waveguide deformation from mechanical motion. The designed coupling gap in ON state is 150 nm, which allows for precise positioning through the mechanical stopper, and simultaneously prevents the waveguides from stiction caused by Van der Waals force due to structural overshooting during mechanical movement. When switching to the second stable state, the air gap in OFF state is more than 1000 nm, sufficient to achieve an ultra-high extinction ratio.

To realize a compact HAC, based on the optimization results of FAQUAD [15–18], we adopted an empirical expression between the width ($w$) and coordinate ($x$) of the HAC:

$$x(w) = \frac{L}{2}\left(c + b\sum_{i=1}^{n} a_i \left(1 + \exp\left(\frac{2\left(\frac{w-w_1}{w_2-w_1}\right)-1}{k_i}\right)\right)^{-1}\right); \sum_{i=1}^{n} a_i = 1 \quad (2)$$

where the waveguide width $w_1$ and $w_2$ are shown in Fig. 2(a), $L$ is the waveguide length, $a_i$ and $k_i$ are the optimization parameters. Other parameters $b$ and $c$ can be calculated through:

$$x(w_1) = 0; \; x(w_2) = L; \quad (3)$$

Compared to the FAQUAD approach, the empirical expression above can be used to directly generate the HAC shape and obtain the desired transmission results in a specific length and bandwidth range. Moreover, $n=2$ can already provide a desirable approximation, where only three parameters ($a_1$, $k_1$ and $k_2$) need to be optimized, thus simplifying the design of the HAC. We used the PSO method in three-dimensional finite-difference time-domain (3D FDTD) simulation to obtain an HAC shape that minimizes crosstalk in ON state, as shown in Fig. 2(b). As an optimization result, in this design, $n=1$, $k_1= 0.11$, $w_1= 0.45$ μm, $w_2= 0.4$ μm, $L = 70$ μm. Fig. 2(c) presents the transmission spectra obtained from ON state simulation, with <0.06 dB excess loss and <−20 dB crosstalk in the 1520-1578 nm bandwidth. Meanwhile, Fig. 2(d) displays the transmission spectra obtained from OFF state simulation, with <0.02 dB excess loss and <−55 dB crosstalk in the same bandwidth.

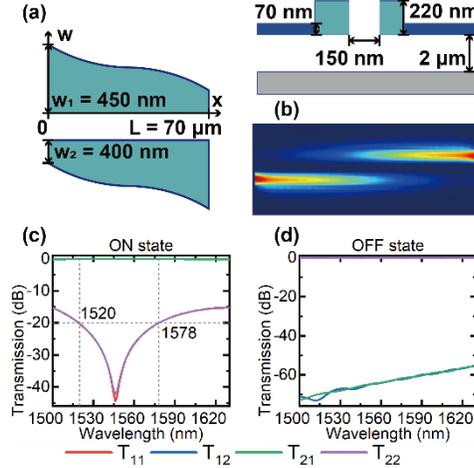

Fig. 2. (a) Width ($w$) and coordinate ($x$) relationship of the designed HAC. The width of the wide waveguide ends is designed to be $w_1 = 450$ nm, while the width of the narrow waveguide ends is designed to be $w_2 = 400$ nm. (b) Simulated light propagation obtained by 3D FDTD. (c) Transmission spectra of ON state. It shows a minimum crosstalk at 1546 nm. (d) Transmission spectra of OFF state.

*2.4 Holistic simulation verification of the entire mechanical structure*

The HAC must be connected to input and output waveguides through suspended waveguides, which are equivalent to mechanical springs in parallel with CCS beams. But using short straight waveguides will result in a high spring constant, preventing the CCS beams to maintain the second stable state, and the photonic switch loses its nonvolatility. Therefore, we use long meandering waveguides to connect the HAC to input and output waveguides, with a minimum radius of curvature designed to be 5 μm, a width of 0.475 μm and a total length of 109.6 μm for each meandering waveguide.

The design of the entire MEMS structure has been verified through FEM simulations. Simulation results show that accelerations up to $1.1 \times 10^5$ $g$ in $y$-direction do not lead to inadvertent switching, indicating excellent mechanical shock resistance. Misalignment of the mechanical stopper with the suspended structure may lead to switch failure. The acceleration required to produce a 220 nm z-direction misalignment obtained from the simulation is $4.6 \times 10^5$ $g$. The maximum tensile and compressive principal strain during switching is estimated to be $\pm 3.2 \times 10^{-3}$, which is significantly below the damage threshold of silicon [19]. The resonant frequencies in ON and OFF states are 72.2 kHz and 58.0 kHz for the fundamental mechanical modes in z-direction, respectively, higher than the mechanical vibration frequencies in most applications, such as datacenters, drones and vehicles. Since there are two sets of CCS beams in each of the proposed nonvolatile switch, the theoretical minimum switching energy for one switch should be 65 fJ and 46 fJ for switching to the second and first stable states, respectively, taking into account the mechanical potential energy of the meandering suspended waveguides.

## 3. Results and discussion

### 3.1 Fabrication

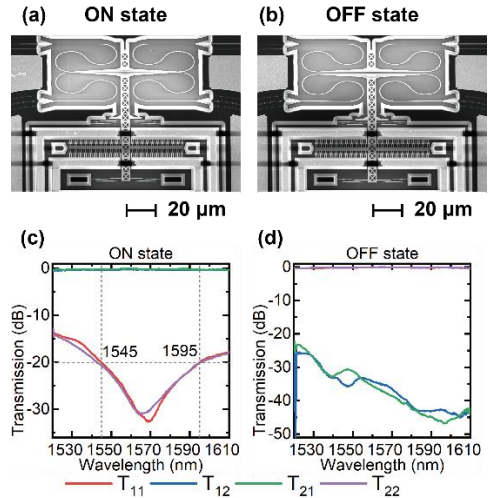

Fig. 3. (a) SEM image of the switch in ON state. (b) SEM image of the switch in OFF state. (c) Transmission spectra of a representative device in ON state. (d) Transmission spectra of a representative device in OFF state.

The nonvolatile silicon photonic MEMS switches are fabricated on a commercial SOI wafer (SOITEC) with a 220-nm-thick top silicon layer and a 2-μm-thick BOX layer. Initially, the ridge waveguides and strip waveguides were patterned using electron beam lithography (EBL), followed by 150 nm shallow etching and 220 nm full etching, respectively. Subsequently, the metal electrodes (50-nm-thick chromium and 300-nm-thick gold films) were patterned using photolithography, followed by an electron beam evaporation and lift-off process. Finally, the suspended structure of the device was released from the BOX by hydrofluoric vapor etching.

Fig. 3(a) and (b) present scanning electron microscope (SEM) images of the switch in ON and OFF states, respectively, with all structures suspended except for the fixed electrostatic combs, the mechanical stoppers and the anchors. The total footprint, including the MEMS actuator, is $100 \times 100$ μm$^2$. In OFF state, the coupling gap of the HAC is >1.1 μm. When the switch is turned on, the electrostatic comb actuator pushes the movable waveguide towards the fixed waveguide and defines the air gap precisely by the mechanical stopper.

## 3.2 Optical performance

Experimental results show that the designed photonic switch has desirable optical performance in C-band. One representative device has been measured with excellent photonic performance across a large bandwidth, as shown in Table 1 and Fig. 3. Fig. 3(c) displays the transmission spectra of the representative device measured in ON state, with crosstalk of –13.93 - –32.65 dB, –12.94 - –31.00 dB for $T_{11}$, $T_{22}$ and excess loss of <0.49 dB, <0.34 dB for $T_{12}$, $T_{21}$ in the measured wavelength range of 1520-1610 nm. Fig. 3(d) shows the transmission spectra of the representative device measured in OFF state, with crosstalk of <–23.31 dB, <–22.15 dB for $T_{12}$, $T_{21}$ and excess loss of <0.43 dB, <0.42 dB for $T_{11}$, $T_{22}$ in the same wavelength range. The representative device has a minimum crosstalk of –32.65 dB (at 1569 nm), –31.00 dB (at 1564 nm) in ON state for $T_{11}$, $T_{22}$.

**Table 1. Optical Performance**

| Switch | Representative Device | |
|---|---|---|
| State | ON | OFF |
| $T_{11}$ (dB) | –13.93 - –32.65 | >–0.43 |
| $T_{12}$ (dB) | >–0.49 | <–23.31 |
| $T_{21}$ (dB) | >–0.34 | <–22.15 |
| $T_{22}$ (dB) | –12.94 - –31.00 | >–0.42 |
| Bandwidth (nm) | 1520-1610 | 1520-1610 |

The optical performance of this switch depends on the overlay error between the two EBL processes, which may result in width errors of the HAC waveguides. Another switch device with the same design parameters as the representative device except 20 nm overlay error compensation shows a higher crosstalk of ~ –5.41 - –6.95 dB in ON state in the measured wavelength range of 1520-1610 nm.

## 3.3 Electro-mechanical performance

The photonic switch exhibits excellent mechanical bistability. A periodic pulsed drive voltage is used to characterize the electro-mechanical properties of the switch. The schematic illustration of the measurement setup is shown in Fig. 4(a). The relationship between the amplitude and minimum required pulse width of the drive pulse from ON to OFF state (switching-OFF pulse) and from OFF to ON state (switching-ON pulse) have been investigated first, as shown in Fig. 4(b). The measured minimal switching voltage (12 V for switching-OFF pulse and 9 V for switching-ON pulse) may suggest a transient out-of-plane switching process because the in-plane switching process should require 21 V for switching-OFF pulse and 20 V for switching-ON pulse according to the simulation results in Fig. 1(e) and the electrostatic force calculations in Section 2.2. As the drive voltage increases, the required pulse width gradually decreases. In the subsequent investigations, we use the switch device with the same design parameters as the representative device except 20 nm overlay error compensation. The pulse width is set to be 5 μs to ensure switching at high frequencies (21 V for switching-OFF pulse and 16 V for switching-ON pulse). For the switching time shown in Fig. 4(c) and (d), switching to ON and OFF states requires 12.9 μs and 3.2 μs, respectively. Periodic pulsed voltage measurements show that our switch can operate up to 20 kHz over $5\times10^7$ switching cycles without observable performance degradation (transmission changes <0.5 dB in both ON and OFF states). After destructive durability test, although there is no observable damage to the switch structure when viewed under SEM, the switch is no longer capable of switching properly. Further investigations on the failure mechanism will be essential to enhance the durability.

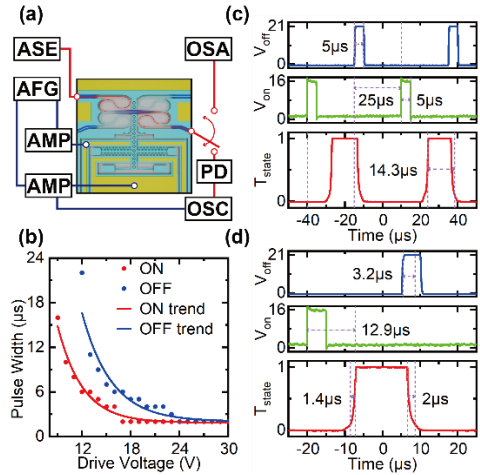

Fig. 4. (a) Schematic illustration of the measurement setup. The blue lines indicate electrical connections, while the red lines indicate optical connections. ASE: Amplified spontaneous emission (light source). PD: Photodetector. OSA: Optical spectrum analyzer. AFG: Arbitrary function generator. AMP: Voltage amplifier. OSC: Oscilloscope. (b) The pulse width and drive voltage relationship of the pulsed signal used. (c) Timing relationship between normalized optical signal ($T_{state}$) and switching voltage signals ($V_{on}$ and $V_{off}$) in a 100-μs time window. $V_{on}$ stands for the switching-ON pulse. $V_{off}$ stands for the switching-OFF pulse. $T_{state}$ shows the ON/OFF state and the switching transients measured by PD. (d) Timing relationship between normalized optical signal and switching voltage signals in a 50-μs time window.

In terms of thermo-mechanical displacement noise within the mechanical structure, the CCS beams in ON state pushes the suspended structure into contact with the mechanical stopper, thereby mechanically fixing the coupling gap of the HAC. Meanwhile, in OFF state, the HAC separates, and the air gap becomes several orders of magnitude larger than the thermo-mechanical displacement of the two HAC waveguides, rendering the effect of displacement noise negligible. Moreover, the temperature change will lead to thermal expansion (or contraction) of the suspended silicon structures, which should enhance (or reduce) the bistability of the CCS beams. Further systematic investigations on the nonvolatility of the switch under varying temperature and other environmental conditions should be conducted before the switch can be employed for real-world applications.

## 4. Conclusion

In this paper, we propose and implement a nonvolatile 2×2 silicon photonic MEMS switch compatible with standard silicon photonic foundry processes. This photonic switch requires 12 V and 9 V drive voltages for switching to OFF and ON states, respectively, and there is no need to maintain the drive voltage after switching is completed. For high optical performance in the limited footprint, we designed a compact HAC with 70-μm length and obtained a simulated extinction ratio of >20 dB in the bandwidth of 1520-1578 nm. One representative device has been measured with excellent photonic performance across a large bandwidth. In the bandwidth of 1520-1610 nm, the switch has an excess loss of <0.49 dB and crosstalk of <–12.94 dB in ON state, and an excess loss of <0.43 dB and crosstalk of <–22.15 dB in OFF state. At 1569 nm, the switch has a minimum crosstalk of –32.65 dB in ON state. The photonic switch has a switching speed in the 10s μs-scale, a footprint of only 100×100 μm$^2$, and a simulated switching energy of 65 fJ and 46 fJ for switching to OFF and ON states, respectively. The photonic switch can operate up to 20 kHz over 5×10$^7$ cycles without observable performance degradation. By properly managing the residual stress and improving the overlay accuracy, it is expected to fabricate photonic switches with better photonic and mechanical performance. The CCS beam design can be adapted for compatibility with standard silicon photonic foundry processes of

130 nm or even 180 nm, providing a highly feasible solution for mass production. This 2×2 switch can be used in a variety of topologies for large-scale photonic switches, and its nonvolatility can potentially support future photonic FPGA designs.

**Funding.** National Science Fund for Distinguished Young Scholars (61725503); National Natural Science Foundation of China (U23B2047, 62321166651, 92150302); Leading Innovative and Entrepreneur Team Introduction Program of Zhejiang (2021R01001); Zhejiang Provincial Major Research and Development Program (2021C01199); Natural Science Foundation of Zhejiang Province (LZ22F050006); Fundamental Research Funds for the Central Universities, and Startup Foundation for Hundred-Talent Program of Zhejiang University.

**Acknowledgments.** The authors thank the ZJU Micro-Nano Fabrication Center and the Westlake Center for Micro/Nano Fabrication and Instrumentation for the facility support.

**Disclosures.** The authors declare no conflict of interest.

**Data availability.** The data that support the findings of this study are available from the corresponding author upon reasonable request.